\documentstyle{elsart}
\begin{document}

\begin{frontmatter}

\noindent INP 1797/PH,  RUB-TPII-6/98 \hfill 15 May 1998

\vspace{0mm}

\title{Tensor susceptibilities of the vacuum from constituent quarks%
\thanksref{grants}}
\thanks[grants]{Research supported by
        the Polish State Committee for
        Scientific Research grant 2P03B-080-12, DFG, BMBF, NATO, 
        and by the Stiftung f\"ur Deutsch-Polnische Zusammenarbeit 
        project 1522/94/LN}
\thanks[emails]{E-mail addresses: broniows@solaris.ifj.edu.pl,
 maximp@hadron.tp2.ruhr-uni-bochum.de, goeke@hadron.tp2.ruhr-uni-bochum.de,
 hchkim@hyowon.cc.pusan.ac.kr}

\author[INP]{Wojciech Broniowski},
\author[RUB]{Maxim Polyakov},
\author[Korea]{Hyun-Chul Kim} and
\author[RUB]{Klaus Goeke}

\address[INP]{H. Niewodnicza\'nski Institute of Nuclear Physics,
         PL-31342 Krak\'ow, Poland}

\address[RUB]{Institute for Theoretical Physics II, 
   Ruhr-Universit\"at-Bochum,
         D-44780 Bochum, Germany}

\address[Korea]{Pusan National University, Dept. of Physics,
    Pusan 609-735, Korea}

\begin{abstract}
We show that the constituent quark model leads to simple expressions for the
isoscalar and isovector tensor
susceptibilities of the vacuum. The found values are negative and
of magnitude compatible  with QCD-sum-rule parameterizations
of spectral densities in
appropriate $L=1$-meson channels.
\end{abstract}

\end{frontmatter}


Susceptibilities of the vacuum are important quantities of strong
interaction physics. They directly enter in the determination of hadron
properties in the QCD sum rule approach. In particular, tensor
susceptibilities of the vacuum \cite{HeJi1,Bel,Kis1} are
relevant for the determination of tensor charges of the nucleon \cite
{HeJi2,JinTang}. Widely different values for these susceptibilities,
all obtained by QCD sum rules techniques, have been reported in the
literature \cite{HeJi1,Bel,Kis1}.
In this paper we address the issue from a completely different viewpoint,
using the concept of {\em constituent quarks}. It is believed that chiral
constituent quark models \cite{mitia:rev,Bijnens} describe properly the
essential physics in the {\em shallow Euclidean} region, with momentum
transfers of the order of $|q^2|\ll\Lambda ^2$, where $\Lambda
$ is the ultraviolet cut off, typically of the order of a few hundred MeV.
This is strongly supported by the instanton-liquid model \cite
{instant1:rev,instant2:rev}. In that region spontaneous breakdown of chiral
symmetry yields a constituent mass for quarks, \mbox{$M\sim (250\div
400){\rm MeV}$}. Susceptibilities are quantities defined with \mbox{$q^2=0$}%
, hence they seem to perfectly fit in the applicability range of the
constituent quark model.

We define the isoscalar and isovector tensor susceptibilities of the vacuum
as
\begin{eqnarray}
\Pi _{I=0} &\equiv &\langle {\rm 0}|i\int d^4x\,T\{\overline{\psi }(x)\sigma
^{\mu \nu }\psi (x),\overline{\psi }(0)\sigma _{\mu \nu }\psi (0)\}|{\rm 0}%
\rangle ,  \label{def0} \\
\Pi _{I=1} &\equiv &\langle {\rm 0}|i\int d^4x\,T\{\overline{\psi }(x)\sigma
^{\mu \nu }\tau ^3\psi (x),\overline{\psi }(0)\sigma _{\mu \nu }\tau ^3\psi
(0)\}|{\rm 0}\rangle ,  \label{def1}
\end{eqnarray}
where $\psi =(u,d)$ is the quark field,
\mbox{$\sigma ^{\mu \nu }=\frac
i2[\gamma^\mu ,\gamma ^\nu ]$}, and \mbox{$\tau^a$} denote Pauli isospin
matrices. Related definitions are present in the literature: He and Ji \cite
{HeJi1} introduce \mbox{$\Pi(0)=\frac{1}{4}(\Pi_{I=0}+\Pi_{I=1})$},
and Belyaev and Oganesian \cite{Bel} consider
\mbox{$\Pi_1(0)\equiv
\frac{1}{12} \Pi(0)$}. The main result of this paper is that the
constituent quark model (CQM)\ predicts:
\begin{equation}
\Pi _{I=0}^{{\rm CQM}}=-24F_\pi^2 \, 
\frac{1+\half \kappa _{I=0}^Q}{g_A^Q}%
,\;\;\;\Pi _{I=1}^{{\rm CQM}}=-24F_\pi^2 \, 
\frac{1+\half \kappa _{I=1}^Q}{g_A^Q},  \label{piquark}
\end{equation}
where \mbox{$F_\pi =93{\rm MeV}$} is the pion decay constant, $g_A^Q$ is the
axial charge of the constituent quark, and \mbox{$\kappa _{I=0,1}^Q$} denote
its isoscalar and isovector anomalous magnetic moments. The result (\ref
{piquark}) is the leading-$N_c$ result, with $g_A^Q$ and $\kappa _{I=0,1}^Q$
entering as model parameters. The expected range for $g_A^Q$ is $%
0.75<g_A^Q\leq 1,$ with the lower value naively helping to reproduce $g_A$
of the nucleon in the non-relativistic quark model. Such lower values are
obtained when constituent quarks attract in the vector-isovector channel
\cite{Bijnens,Klimt,Vogl}. In the absence of such interactions 
$g_A^Q=1$. It is commonly accepted that the magnitudes of $\kappa _{I=0}^Q$
and $\kappa _{I=1}^Q$ are tiny, less than $\sim 5\%$. In constituent quark
models these quantities depart from zero when tensor interactions among quarks
are introduced \cite{Klimt,Vogl}. Using above estimates we find
\begin{equation}
-0.25{\rm GeV}^2\leq \Pi _{I=0,1}^{{\rm CQM}}\leq -0.2{\rm GeV}^2.
\label{piqnum}
\end{equation}
This value is very close to the estimate of Belyaev and Oganesian
\cite{Bel} made in the framework of the QCD sum rules: 
\mbox{$(\Pi_{I=1}+(\Pi _{I=0})/2=-0.2 {\rm GeV}^2$}.

Now we pass to the derivation of Eq.~(\ref{piquark}). For our purpose no
details of a specific constituent quark model are needed, such as the way of
introducing the ultraviolet cut off, its value, or the value of the
constituent quark mass. It is enough that the spontaneous symmetry has been
broken and quarks are massive. In addition, our calculation is made in
the {\em leading-}$N_c${\em \ approximation}, which is equivalent to the 
one-quark-loop approximation in the effective action.
Corrections to this leading result could
in principle be included. We begin with the simplest case of no tensor nor
vector interactions. Then Eqs.~(\ref{def0}-\ref{def1}) are evaluated by the
one-quark-loop diagram, which in the momentum space gives
\begin{eqnarray}
\Pi _{I=0,1}^{{\rm CQM\ }} &=&N_cN_f\,i\int \frac{d^4k}{(2\pi )^4}\,\frac{%
{\rm Tr}_{{\rm Dirac}}[(\FMSlash{k}+M)\sigma ^{\mu \nu }
(\FMSlash{k}+M)\sigma _{\mu \nu }]}{%
(k^2-M^2+i\delta )^2}=  \label{div} \\
\  &=&48N_cN_fi\int \frac{d^4k}{(2\pi )^4}\,\frac{M^2}{(k^2-M^2+i\delta )^2}%
=-48N_cN_f\int_\Lambda \frac{d^4k_E}{(2\pi )^4}\,\frac{M^2}{(k_E^2+M^2)^2},
\nonumber
\end{eqnarray}
where $N_c=3$ is the number of colors, $N_f=2$ is the number of flavors, and
$M$ is the mass of the constituent quark. In the last equality we have
Wick-rotated to Euclidean momentum $k_E$. The integral in Eq.~(\ref{div}) is
divergent, as is inherent in effective models, hence a regulator is
introduced in the integral in Eq.~(\ref{div}), indicated by the subscript $%
\Lambda $ in the integral. There are many ways of introducing the cut-off,
the details, however, are irrelevant. The crucial point is that on the RHS
of Eq.~(\ref{div}) we recognize the well-known result of constituent quark
models \cite{mitia:rev,Bijnens,Klimt,Vogl} for the square of the
pion decay constant:
\begin{equation}
F_\pi ^2=4N_c\int_\Lambda \frac{d^4k_E}{(2\pi )^4}\,\frac{M^2}{%
(k_E^2+M^2)^2}.  \label{fpi}
\end{equation}
Combining Eqs.~(\ref{div}-\ref{fpi}) proves Eq.~(\ref{piquark}) for the case
of no vector-isovector ($g_A^Q=1)$ and no tensor ($\kappa _{I=,01}^Q=0$)
interaction. Note that our result holds also for the case where $M$ depends
on the momentum, $M=M(k^2)$, as in the case of the instanton-vacuum 
model described in Ref.~\cite{mitia:rev}.

In the general case interactions in the vector and tensor channels can be
present. The following terms in the effective chiral SU(2)$\otimes $SU(2)
Lagrangian density are relevant for our purpose:
\begin{eqnarray}
&&-\frac 12G_\rho \left[ \left( \overline{\psi }(x)\gamma ^\mu \tau ^a\psi
(x)\right) ^2+\left( \overline{\psi }(x)\gamma ^\mu \gamma _5\tau ^a\psi
(x)\right) ^2\right]  \nonumber \\
&&+\frac 12G_T\left[ \left( \overline{\psi }(x)\sigma ^{\mu \nu }\psi
(x)\right) ^2-\left( \overline{\psi }(x)\sigma ^{\mu \nu }\tau ^a\psi
(x)\right) ^2\right] .  \label{vecten}
\end{eqnarray}
The signs inside brackets are fixed by the chiral invariance, and the signs
outside brackets are conventional. The presence of interactions (\ref{vecten}%
) causes rescattering of quarks in vector and tensor channels: for instance,
the $q\overline{q}$ pair induced by the tensor current interacts, while
propagating, by the tensor interactions. The resulting Bethe-Salpeter chain
can be summed up as a geometric series. Details of this procedure are well
known \cite{Bijnens} and we give no further details here.
The effects of finite $G_\rho $ and $G_T$ are particularly simple for
the vanishing external momentum, $q=0$, 
which is the case needed here. We find
\begin{equation}
\Pi _{I=0,1}^{{\rm CQM}}=-24\frac{J(0)}{1\pm 4G_TJ(0)},  \label{ten}
\end{equation}
where $J(0)=4N_c \int_\Lambda \frac{d^4k_E}{(2\pi )^4}\,M^2/(k_E^2+M^2)^2.$
Also, the expressions for the pion decay constant and the weak charge of the
constituent quark are modified appropriately 
by the presence of vector-isovector
interactions \cite{Bijnens,Klimt,Vogl},
\begin{equation}
F_\pi ^2=g_A^QJ(0),\quad \quad g_A^Q=\frac 1{1+4G_\rho J(0)}.  \label{gaq}
\end{equation}
The anomalous magnetic moments of the constituent quark are equal to \cite
{Bijnens,Klimt,Vogl}
\begin{equation}
\kappa _{I=0,1}^Q=\frac{\mp 8G_TJ(0)}{1\pm 4G_TJ(0)},\quad {\rm or:\quad }1+%
\frac{\kappa _{I=0,1}^Q}2=\frac 1{1\pm 4G_TJ(0)}.  \label{kappa}
\end{equation}
With the help of relations (\ref{gaq},\ref{kappa}) we rewrite Eq.~(\ref{ten}%
) as Eq.~(\ref{piquark}), completing the proof.

In the remaining part of this paper we argue that the sign and size of $\Pi
_{I=0}$ and $\Pi _{I=1},$ as given by Eq.~(\ref{piqnum}), agree with
expectations based on dispersion relations and simple models of spectra in
appropriate channels. Let us begin with $\Pi _{I=1},$ for which the relevant
intermediate states have quantum numbers of $\rho $ ($%
I^G(J^{PC})=1^{+}(1^{--})$) and $b_1(1235)$ ($I^G(J^{PC})=1^{+}(1^{+-})$).
With the help of the spectral decomposition we can write the imaginary part
of the correlator
\mbox{$\Pi_{I=1}(q)\equiv
i\langle {\rm 0}|\int d^4x\,e^{iq\cdot x}T\{\overline{\psi }(x)\sigma ^{\mu
\nu }\tau^3 \psi (x),\overline{\psi }(0)\sigma _{\mu \nu }\tau^3
\psi (0)\}|{\rm 0}\rangle
$} as
\begin{eqnarray}
\!\!\!\!\!\!\!\!\!\!
{\rm Im}\Pi _{I=1}(q) &=&\pi \sum_\rho \delta (q^2-m_\rho ^2)\sum_\lambda
\langle {\rm 0}|\overline{\psi }(0)\sigma ^{\mu \nu }\tau ^3\psi (0)|\rho
;q,\lambda \rangle \langle \rho ;q,\lambda |\overline{\psi }(0)\sigma _{\mu
\nu }\tau ^3\psi (0)|{\rm 0}\rangle +  \nonumber \\
&+&\pi \sum_b\delta (q^2-m_b^2)\sum_\lambda \langle {\rm 0}|%
\overline{\psi }(0)\sigma ^{\mu \nu }\tau ^3\psi (0)|b;q,\lambda \rangle
\langle b;q,\lambda |\overline{\psi }(0)\sigma _{\mu \nu }\tau ^3\psi (0)|%
{\rm 0}\rangle , \label{impi}
\end{eqnarray}
where $\sum_{\rho ,b}$ denote the sum/integral over all physical states with
quantum numbers of neutral $\rho $ and $b_1,$ quantities $m_\rho $ and $m_b$
denote the invariant mass of these states, and $\lambda $ are their
polarizations. The Lorentz structure of the matrix elements consistent with
quantum numbers of the state is\footnote{The 
factors of $\sqrt{2}$ in Eqs.~(\ref{omel}-\ref{hel}) make our expressions 
consistent with the definition \mbox{$\langle {\rm 0}|%
\overline{u}(0)\sigma ^{\mu \nu }d(0)|\rho ^{+};q,\lambda \rangle =if_\rho
(q^2)\left( \epsilon _\lambda ^\mu q^\nu -\epsilon _\lambda ^\nu q^\mu
\right)$} of Ref.~\cite{Ball}.}
\begin{eqnarray}
\langle {\rm 0}|\overline{\psi }(0)\sigma ^{\mu \nu }\tau ^3\psi (0)|\rho
;q,\lambda \rangle  &=&i\sqrt{2}f_\rho (q^2)\left( \epsilon _\lambda ^\mu
q^\nu -\epsilon _\lambda ^\nu q^\mu \right) ,  \label{omel} \\
\langle {\rm 0}|\overline{\psi }(0)\sigma ^{\mu \nu }\tau ^3\psi
(0)|b;q,\lambda \rangle  &=&i\sqrt{2}f_b(q^2)\varepsilon ^{\mu \nu \alpha
\beta }(\epsilon _\lambda )_\alpha q_\beta ,  \label{hel}
\end{eqnarray}
where $\epsilon _\lambda $ is the polarization vector. 
Using the identity $\sum_\lambda
\epsilon _\lambda ^\mu (\epsilon _\lambda ^\nu )^{*}=-g^{\mu \nu }+q^\mu
q^\nu /q^2$ we find that
\begin{eqnarray}
\sum_\lambda \langle 0|\overline{\psi }(0)\sigma ^{\mu \nu }\tau ^3\psi
(0)|\rho ;q,\lambda \rangle \langle \rho ;q,\lambda |\overline{\psi }%
(0)\sigma _{\mu \nu }\tau ^3\psi (0)|0\rangle  &=&-12q^2|f_\rho (q^2)|^2,
\label{A} \\
\sum_\lambda \langle 0|\overline{\psi }(0)\sigma ^{\mu \nu }\tau ^3\psi
(0)|b;q,\lambda \rangle \langle b;q,\lambda |\overline{\psi }(0)\sigma _{\mu
\nu }\tau ^3\psi (0)|0\rangle  &=&+12q^2|f_b(q^2)|^2,  \label{B}
\end{eqnarray}
and, consequently, Eq.~(\ref{impi}) can be written as
\begin{equation}
{\rm Im}\Pi _{I=1}(q)=-12\pi \sum_\rho \delta (q^2-m_\rho ^2)m_\rho
^2|f_\rho (m_\rho ^2)|^2+12\pi \sum_b\delta (q^2-m_b^2)m_b^2|f_b(m_b^2)|^2.
\label{impis}
\end{equation}
Note the negative sign of the $\rho $ contribution, and the positive sign of
the $b_1$ contribution. The dispersion relation gives
\begin{equation}
\Pi _{I=1}(q)=\frac{12}\pi \int_0^\infty ds\frac{\sigma _b(s)-\sigma _\rho
(s)}{s-q^2-i\delta},  \label{disp}
\end{equation}
where in general the spectral densities collect the contribution from all
poles (and cuts) in the appropriate channel: $\sigma (s)=\sum_a\delta
(s-m_a^2)m_a^2|f_a(m_a^2)|^2$. Dispersion relation (\ref{disp}) requires no
subtraction in the chiral limit. This follows from the fact that in QCD the
asymptotic forms at large Euclidean momenta ($q^2\rightarrow -\infty $) of
tensor correlators in the $\rho $ and $b_1$ channels are equal \cite{GL1}.
Indeed, the tensor correlator
\begin{equation}
\Pi _{I=1}^{\mu \nu ,\alpha \beta }(q)\equiv i\langle {\rm 0}|\int
d^4x\,e^{iq\cdot x}T\{\overline{\psi }(x)\sigma ^{\mu \nu }\tau ^3\psi (x),%
\overline{\psi }(0)\sigma ^{\alpha \beta }\tau ^3\psi (0)\}|{\rm 0}\rangle
\label{tenq}
\end{equation}
can be decomposed as as follows in structures involving the contributions of
$\rho $ and $b_1$ states:
\begin{eqnarray}
\Pi _{I=1}^{\mu \nu ,\alpha \beta }(q) &=&-4T_{(-)}^{\mu \nu ,\alpha \beta
}\Pi _\rho (q)+4T_{(+)}^{\mu \nu ,\alpha \beta }\Pi _b(q),  \label{tendec}
\end{eqnarray}
where the factors of $4$ are conventional\footnote{%
We note that the constituent quark model gives
$\Pi^{\rm CQM}_\rho(0)+\Pi^{\rm CQM}_b(0)=0$, in agreement with the general
requirement due to the absence of zero-mass states in the tensor channel
\cite{Ball,Cher}.}. The tensor structures are defined
as
\begin{eqnarray}
T_{(-)}^{\mu \nu ,\alpha \beta \ } &=&\frac 1{2q^2}\left( q^\alpha q^\mu
g^{\beta \nu }-q^\beta q^\mu g^{\alpha \nu }+q^\beta q^\nu g^{\alpha \mu
}-q^\alpha q^\nu g^{\beta \mu }\right) ,  \label{tenrho} \\
T_{(+)}^{\mu \nu ,\alpha \beta \ } &=&-T_{(-)}^{\mu \nu ,\alpha \beta \
}+\frac 12\left( g^{\alpha \mu }g^{\beta \nu }-g^{\alpha \nu }g^{\beta \mu
}\right) ,  \label{tenb}
\end{eqnarray}
and satisfy $T_{(-)\mu \nu }^{\mu \nu}=T_{(+)\mu \nu }^{\mu \nu}=3$.
In the deep Euclidean region QCD gives \cite{Ball,GL1}\footnote{%
Ref. \cite{GL1} has an error in the overall factor in Eq.~(A3). Their
tensors in Eqs.~(2.3) relate to ours as follows:
 $P^{(1)}=\frac{1}{6}T_{(-)}$, $%
 P^{(2)}=-\frac{1}{6}T_{(+)}$, and Eq.~(A3) should read: \mbox{$\Pi
(q)=24P^{(1)}\Pi^{-}+24P^{(1)}\Pi^{+}$}, in order to agree with our Eq.~(\ref
{tendec})}.
\begin{equation}
\Pi_{\rho ,b}^{{\rm QCD}}(q) =-\frac{q^2}{8\pi ^2}\left\{ \left( 1\pm
\frac{6m^2}{q^2}\right) \log \frac{-q^2}{\mu ^2} +%
{\cal O}(\alpha_s)\right\} +{\cal O}(\frac 1{q^2}),  \label{asympt}
\end{equation}
where $m$ is the current quark mass, which is neglected. The form (\ref
{asympt}) implies that at large $s$ we have asymptotically 
\mbox{$\sigma_\rho(s) \sim \sigma_b(s)\sim  s /(8\pi)$}, and the
difference \mbox{$\sigma _\rho (s)-\sigma _b(s)\sim 1/s$}. 
This means that the
unsubtracted dispersion relation (\ref{disp}) holds, and we obtain
immediately
\begin{equation}
\Pi_{I=1}=\frac{12}\pi \int_0^\infty ds\frac{\sigma_b(s)-\sigma_\rho (s)}%
s.  \label{disp1}
\end{equation}
Repeating the above analysis for the isovector channel, where the relevant
physical states have the quantum numbers of $\omega $ ($%
I^G(J^{PC})=0^{-}(1^{--})$) and $h_1(1170)$ ($I^G(J^{PC})=0^{-}(1^{+-})$),
with conventionally the same factors as in Eqs.~(\ref{omel},\ref{hel},%
\ref{tendec},\ref{asympt}) gives
\begin{equation}
\Pi _{I=0}=\frac{12}\pi \int_0^\infty ds\frac{\sigma _h(s)-\sigma _\omega (s)%
}s.  \label{disp2}
\end{equation}
Note that in Eqs.~(\ref{disp1}-\ref{disp2}) there is cancelation between
spectral densities corresponding to opposite parity states.

Now, following Refs. \cite{HeJi1,Bel,Ball}, we can
estimate the contributions to
Eq.~(\ref{disp1}-\ref{disp2}). We expect that the following pole + continuum
parameterization of the spectral densities should give at least a correct
order-of-magnitude:
\begin{equation}
\sigma _a(s)=\pi m_a^2{\rm f}_a^2\delta (s-m_a^2)+\frac s{8\pi }
\theta (s_a-s),  \label{modpar}
\end{equation}
where $a$ labels the channel. 
Substituting Eq.~(\ref{modpar}) into Eqs.~(\ref
{disp1}-\ref{disp2}) gives
\begin{eqnarray}
\Pi _{I=0} &=&-12({\rm f}_\omega ^2-{\rm f}_h^2+\frac 1{8\pi
^2}(s_h-s_\omega )),\quad   \label{disp1mod} \\
\Pi _{I=1} &=&-12({\rm f}_\rho ^2-{\rm f}_b^2+\frac 1{8\pi ^2}(s_b-s_\rho ))
\label{disp2mod}
\end{eqnarray}
Let us begin with the last term in Eq.~(\ref{disp2mod}). The $s_b$ parameter
has been obtained from QCD sum rules in Refs. \cite{Bel,Ball,GL1} giving $%
s_b\sim (2.3\div 3){\rm GeV}^2$. Using derivative coupling Ref.~\cite{Ovch}
reports $s_b\sim 2.5{\rm GeV}^2.$ The QCD sum rule fit for the $\rho$ meson
spectrum in the tensor channel gives \cite{Ball,Cher,Bak} $s_\rho \sim
(1.3\div 1.7){\rm GeV}^2$. Thus we have $s_b-s_\rho \sim 0.6\div 1.7{\rm GeV}%
^2,$ and the continuum contribution to $\Pi_{I=1}$ is of the order of $-0.1%
{\rm GeV}^2$ to $-0.25{\rm GeV}^2$. The $b_1$ pole contribution is,
according to Ref.~\cite{Bel,Ball},
$12{\rm f}_b^2\sim 0.4{\rm GeV}^2$, and the $\rho$ pole
contribution is, according to Ref. \cite{Ball}, $12{\rm f}_\rho ^2\sim 0.3%
{\rm GeV}^2,$ according to \cite{Bak}, $12{\rm f}_\rho ^2\sim 0.35{\rm GeV}%
^2$, and according to \cite{Cher}, $12{\rm f}_\rho ^2\sim 0.5{\rm GeV}^2$.
Taking a ``global'' average yields $\Pi _{I=1}\sim (-0.1\div -0.25){\rm GeV}%
^2$. Note that the order of magnitude of all contributions to Eq.~(\ref
{disp2mod}) is the same as in Eq.~(\ref{piqnum}), indicating our
constituent-quark-model value has the expected size. The sign of Eq.~(\ref
{piqnum}) shows that the $\rho $ channel ``wins'' over the $b_1$ channel in
Eq.~(\ref{disp2mod}). In fact, as already pointed out in Ref.~\cite{HeJi1},
cancelations in Eqs.~(\ref{disp1mod}-\ref{disp2mod}) make any precise
estimates of tensor susceptibilities based on models of spectra difficult if
not impossible. Note, however,
as pointed out by Belyaev and Oganesian \cite{Bel},
that Ref. \cite{HeJi1} does not include the continuum contribution in Eq.~(%
\ref{disp2mod}), which causes their value to be much smaller, and of
opposite sign compared to Ref.~\cite{Bel} and
to our estimate (\ref{piqnum}). The
value for $\Pi(0)$ obtained by Kisslinger \cite
{Kis1} from the QCD sum rules for three-point functions are of opposite sign
and large in magnitude. In view of these controversies, direct model
estimates, such as that of Eq.~(\ref{piqnum}) gain significance.

We point out that the mentioned QCD sum rule calculations do not distinguish
between the isoscalar and isovector channels. Surprisingly, it
is supported by or result
(\ref{piquark}), since, due to the smallness of the quark anomalous magnetic
moments, $\Pi_{I=0}^{{\rm CQM}}$ is practically equal to $\Pi_{I=1}^{{\rm %
CQM}}$. 
A corollary of this observation is the approximate relation\\
\mbox{$\int_0^\infty ds (\sigma _h-\sigma _\omega )/s \simeq 
\int_0^\infty ds (\sigma _b-\sigma _\rho)/s$}.

We conclude with several comments:
\begin{enumerate}
\item  The obtained values for tensor susceptibilities of the vacuum have
direct relevance for the calculations of tensor charges of the nucleon \cite
{HeJi1,HeJi2,JinTang}.
\item  The simple result (\ref{piquark}) and the following numerical
estimate (\ref{piqnum}) are leading-$N_c$ results. Formally suppressed
effects, such as meson (especially pion) exchange, may be important, as is the
case {\em e.g.} in the calculation of the quark condensate in the
Nambu--Jona-Lasinio model \cite{Emil}.
\item  Quark-model relations, such as our Eq.~(\ref{piquark}), bear some
similarity to the relations obtained by Leutwyler \cite{Leut} and Mallik \cite
{Mallik} on the basis of the SU(6) symmetry of meson wave functions. In
fact, more relations of the type of Eq.~(\ref{piquark}) can be obtained in
our approach \cite{tobe}.
\item  Our relations of susceptibilities to the $L=1$ meson spectra, Eqs.(%
\ref{disp1}-\ref{disp2}), are obtained via dispersion relations. This is in
the sprit of QCD sum rules, where one compares the deep Euclidean region
to the physical region via dispersion relations. In our method we compare
the shallow Euclidean region to the physical region, following the idea
described in Ref. \cite{Boch}. That way we can infer useful information
concerning meson excitations without leaving the validity range of the
constituent chiral quark model.
\end{enumerate}
We thank Rusko Ruskov for many discussions on vacuum susceptibilities in QCD
sum rules.
%
%

\end{document}